\documentclass[twocolumn,times]{aastex63}
\usepackage{graphicx}
\usepackage{makecell}
\usepackage{amsmath}
\usepackage{amsthm}
\usepackage{booktabs}
\usepackage{lineno}
\usepackage{threeparttable}
\usepackage[squaren]{SIunits}
\usepackage{xspace}
\usepackage{comment}
\usepackage{xcolor}
\usepackage{multirow}

\usepackage{array}

\setwatermarkfontsize{1.5in}

\shortauthors{Xiao ET AL.}
\begin{document}

\title{Search for Quasi-Periodical Oscillations in Precursors of Short and Long Gamma Ray Bursts}
\correspondingauthor{Wen-Xi Peng, Shuang-Nan Zhang, Shao-Lin Xiong}
\email{pengwx@ihep.ac.cn, zhangsn@ihep.ac.cn, xiongsl@ihep.ac.cn}
\author{Shuo Xiao}
\affil{Guizhou Provincial Key Laboratory of Radio Astronomy and Data Processing, Guizhou Normal University, Guiyang 550001, People’s Republic of China}
\affil{School of Physics and Electronic Science, Guizhou Normal University, Guiyang 550001, People’s Republic of China}

\author{Wen-Xi Peng*}
\affil{Key Laboratory of Particle Astrophysics, Institute of High Energy Physics, Chinese Academy of Sciences, Beijing 100049, China}

\author{Shuang-Nan Zhang*}
\affil{Key Laboratory of Particle Astrophysics, Institute of High Energy Physics, Chinese Academy of Sciences, Beijing 100049, China}

\author{Shao-Lin Xiong*}
\affil{Key Laboratory of Particle Astrophysics, Institute of High Energy Physics, Chinese Academy of Sciences, Beijing 100049, China}

\author{Xiao-Bo Li}
\affil{Key Laboratory of Particle Astrophysics, Institute of High Energy Physics, Chinese Academy of Sciences, Beijing 100049, China}

\author{You-Li Tuo}
\affil{Key Laboratory of Particle Astrophysics, Institute of High Energy Physics, Chinese Academy of Sciences, Beijing 100049, China}

\author{He Gao}
\affil{Department of Astronomy, Beijing Normal University, Beijing 100875, People’s Republic of China}

\author{Yue Wang}
\affil{Key Laboratory of Particle Astrophysics, Institute of High Energy Physics, Chinese Academy of Sciences, Beijing 100049, China}
\affil{University of Chinese Academy of Sciences, Chinese Academy of Sciences, Beijing 100049, China}

\author{Wang-Chen Xue}
\affil{Key Laboratory of Particle Astrophysics, Institute of High Energy Physics, Chinese Academy of Sciences, Beijing 100049, China}
\affil{University of Chinese Academy of Sciences, Chinese Academy of Sciences, Beijing 100049, China}

\author{Chao Zheng}
\affil{Key Laboratory of Particle Astrophysics, Institute of High Energy Physics, Chinese Academy of Sciences, Beijing 100049, China}
\affil{University of Chinese Academy of Sciences, Chinese Academy of Sciences, Beijing 100049, China}

\author{Yan-Qiu Zhang}
\affil{Key Laboratory of Particle Astrophysics, Institute of High Energy Physics, Chinese Academy of Sciences, Beijing 100049, China}
\affil{University of Chinese Academy of Sciences, Chinese Academy of Sciences, Beijing 100049, China}

\author{Jia-Cong Liu}
\affil{Key Laboratory of Particle Astrophysics, Institute of High Energy Physics, Chinese Academy of Sciences, Beijing 100049, China}
\affil{University of Chinese Academy of Sciences, Chinese Academy of Sciences, Beijing 100049, China}

\author{Cheng-Kui Li}
\affil{Key Laboratory of Particle Astrophysics, Institute of High Energy Physics, Chinese Academy of Sciences, Beijing 100049, China}

\author{Shu-Xu Yi}
\affil{Key Laboratory of Particle Astrophysics, Institute of High Energy Physics, Chinese Academy of Sciences, Beijing 100049, China}

\author{Xi-Lu Wang}
\affil{Key Laboratory of Particle Astrophysics, Institute of High Energy Physics, Chinese Academy of Sciences, Beijing 100049, China}

\author{Zhen Zhang}
\affil{Key Laboratory of Particle Astrophysics, Institute of High Energy Physics, Chinese Academy of Sciences, Beijing 100049, China}

\author{Ce Cai}
\affil{College of Physics, Hebei Normal University, 20 South Erhuan Road, Shijiazhuang, 050024, China}

\author{Ai-Jun Dong}
\affil{Guizhou Provincial Key Laboratory of Radio Astronomy and Data Processing, Guizhou Normal University, Guiyang 550001, People’s Republic of China}
\affil{School of Physics and Electronic Science, Guizhou Normal University, Guiyang 550001, People’s Republic of China}

\author{Wei Xie}
\affil{Guizhou Provincial Key Laboratory of Radio Astronomy and Data Processing, Guizhou Normal University, Guiyang 550001, People’s Republic of China}
\affil{School of Physics and Electronic Science, Guizhou Normal University, Guiyang 550001, People’s Republic of China}

\author{Jian-Chao Feng}
\affil{Guizhou Provincial Key Laboratory of Radio Astronomy and Data Processing, Guizhou Normal University, Guiyang 550001, People’s Republic of China}
\affil{School of Physics and Electronic Science, Guizhou Normal University, Guiyang 550001, People’s Republic of China}

\author{Qing-Bo Ma}
\affil{Guizhou Provincial Key Laboratory of Radio Astronomy and Data Processing, Guizhou Normal University, Guiyang 550001, People’s Republic of China}
\affil{School of Physics and Electronic Science, Guizhou Normal University, Guiyang 550001, People’s Republic of China}

\author{De-Hua Wang}
\affil{Guizhou Provincial Key Laboratory of Radio Astronomy and Data Processing, Guizhou Normal University, Guiyang 550001, People’s Republic of China}
\affil{School of Physics and Electronic Science, Guizhou Normal University, Guiyang 550001, People’s Republic of China}

\author{Xi-Hong Luo}
\affil{Guizhou Provincial Key Laboratory of Radio Astronomy and Data Processing, Guizhou Normal University, Guiyang 550001, People’s Republic of China}
\affil{School of Physics and Electronic Science, Guizhou Normal University, Guiyang 550001, People’s Republic of China}

\author{Qi-Jun Zhi}
\affil{Guizhou Provincial Key Laboratory of Radio Astronomy and Data Processing, Guizhou Normal University, Guiyang 550001, People’s Republic of China}
\affil{School of Physics and Electronic Science, Guizhou Normal University, Guiyang 550001, People’s Republic of China}

\author{Li-Ming Song}
\affil{Key Laboratory of Particle Astrophysics, Institute of High Energy Physics, Chinese Academy of Sciences, Beijing 100049, China}

\author{Ti-Pei Li}
\affil{Key Laboratory of Particle Astrophysics, Institute of High Energy Physics, Chinese Academy of Sciences, Beijing 100049, China}
\affil{Department of Astronomy, Tsinghua University, Beijing 100084, People’s Republic of China}

\begin{abstract}
The precursors of short and long Gamma Ray Bursts (SGRBs and LGRBs) can serve as probes of their progenitors, as well as shedding light on the physical processes of mergers or core-collapse supernovae. Some models predict the possible existence of Quasi-Periodically Oscillations (QPO) in the precursors of SGRBs. Although many previous studies have performed QPO search in the main emission of SGRBs and LGRBs, so far there was no systematic QPO search in their precursors. In this work, we perform a detailed QPO search in the precursors of SGRBs and LGRBs detected by Fermi/GBM from 2008 to 2019 using the power density spectrum (PDS) in frequency domain and Gaussian processes (GP) in time domain. We do not find any convinced QPO signal with significance above 3 $\sigma$, possibly due to the low fluxes of precursors. Finally, the PDS continuum properties of both the precursors and main emissions are also studied for the first time, and no significant difference is found in the distributions of the PDS slope for precursors and main emissions in both SGRBs and LGRBs.

\end{abstract}


\section{Introduction}
Gamma-Ray Bursts (GRBs) are traditionally divided into long GRBs (LGRBs) and short GRBs (SGRBs) based on their duration (separated at about 2 s). LGRBs are widely believed to origin from core collapses of massive stars and associated with supernovae, whereas SGRBs are related to binary neutron star or neutron star-black hole mergers (\citealp{woosley2006supernova}; \citealp{abbott2017gravitational}) where a kilonova due to a series of r-process nucleosynthesis \citep{li1998transient,wu2017imprints} can be seen. Precursors are detected in about 10\% of GRBs detected by Swift and most of them are LGRBs, and no significant difference was found between the precursor emission and the main emission episodes, suggesting that the precursors are directly related to the central engine activities and thus share the same physical origin with the prompt episode \citep{hu2014internal,li2022temporal}. A few SGRBs, about 2.7\% of SGRBs detected by Swift/BAT and Fermi/GBM, also have precursor emission, which usually have a blackbody or exponential cutoff power-law (CPL) spectra (\citealp{zhong2019precursors}; \citealp{wang2020stringent}). The thermal precursor of SGRB may originate from the shock breakout or the photospheric radiation of a fireball launched after the merger, whereas nonthermal precursors may be explained by magnetospheric interaction between two neutron stars prior to the merger (\citealp{hansen2001radio,troja2010precursors}; \citealp{palenzuela2013electromagnetic}; \citealp{wang2018pre}), or the the resonant shattering of the crusts of neutron star \citep{tsang2012resonant}.

Many previous studies have performed QPO search in main emission of SGRBs (e.g \citealp{kruger2002search,zhilyaev2009detection,dichiara2013search}) and LGRBs (e.g. \citealp{kruger2002search,cenko2010unveiling,guidorzi2016individual,tarnopolski2021comprehensive}), where the former may originate from quasi-periodic jet precession from black hole (BH)–neutron star (NS) mergers \citep{stone2013pulsations}, and the latter may be related to the long activity of the magnetars \citep{2009GCN..9645....1M}. However, none of the claimed QPOs has been confirmed by further studies. 

On the other hand, some models predict the possible existence of QPOs in precursors of SGRBs, such as pre-merger magnetosphere interaction \citep{xiao2022quasi}, magnetar tidal-induced super flare model \citep{zhang2022tidal}, seismic aftershocks from resonant scattering of neutron star \citep{sarin2021evolution,suvorov2022quasi} and the accelerating relativistic binary winds via the synchrotron maser process \citep{sridhar2021shock}. Indeed, recently a Quasi-Periodic
oscillations (QPOs) has been reported in the non-thermal precursor of GRB 211211A \citep{xiao2022quasi}. This GRB lasts for about 50 seconds \citep{mangan2021grb}; however, a kilonova was detected \citep{rastinejad2022kilonova,xiao2022quasi} accompanying the GRB, and its spectral lag is only several milliseconds, consistent with that of SGRB \citep{xiao2022robust,troja2022long}. Besides, its positions on the spectral Lag - luminosity ($Lag - L_{\rm iso}$) and Amati diagrams are consistent with the merger origin (\citealp{xiao2022quasi,rastinejad2022kilonova}; \citealp{yang2022peculiarly}). This QPO may be generated by the torsional or crustal oscillations of the magnetar (\citealp{xiao2022quasi}; \citealp{gao2022grb}; \citealp{2022arXiv220511112S}; \citealp{zhang2022tidal}). 

However, so far there was no systematic QPO search in the precursors of both SGRBs and LGRBs, which is the focus of this work. We present our sample selection and search methods in Section 2, and give the results in Section 3. Finally, discussion and summary are given in Section 4.
%加上长暴precursor和主暴一样可能存在qpo

\section{SAMPLE SELECTION and Methodology}

\begin{table*}[htbp]
\caption{\centering QPOs search for precursors in SGRBs using Power density spectrum and Gaussian processes methods.}\label{SGRBs_table}
\begin{tabular*}{\hsize}{@{}@{\extracolsep{\fill}}ccccccccc@{}}
	\hline
Name$^a$         & Tpre (s) & Li-Ma Significance$^b$ ($\sigma$)& $f{\rm _{PDS}}$ (Hz) &$p_{\rm value}({\rm PDS})$ & $f{\rm _{GP}}$ (Hz)  &  ${\rm In}BF_{\rm qpo}({\rm GP})$$^c$  \\
	\hline
GRB 081024245  & 0.06 & 1.9  & 556 & 0.98 & 475 & -1.9 \\
GRB 081216531  & 0.15 & 4.1  & 156 & 0.86 & 174 & 0.3  \\
GRB 090510016  & 0.05 & 5.7  & 621 & 0.53 & 119 & 0.2  \\
GRB 100223110 & 0.02 & 3.5  & 226 & 0.74 & 226 & 0.2  \\
GRB 100717372 & 0.15 & 7.6  & 38  & 0.79 & 301 & -1.0 \\
GRB 100827455 & 0.11 & 3.7  & 264 & 0.1  & 144 & -0.5 \\
GRB 101208498 & 0.17 & 7.1  & 939 & 0.7  & 429 & -0.2 \\
GRB 111117510 & 0.18 & 13.7 & 358 & 0.81 & 287 & -0.2 \\
GRB 130310840 & 0.58 & 2.6  & 458 & 0.06 & 459 & -3.2 \\
GRB 140209313 & 0.61 & 7.1  & 53  & 0.29 & 8   & 1.0  \\
GRB 141102536 & 0.06 & 5.0  & 362 & 0.32 & 302 & 0.9  \\
GRB 150604434 & 0.17 & 9.5  & 884 & 0.6  & 499 & -0.4 \\
GRB 150922234 & 0.05 & 9.2  & 709 & 0.56 & 341 & -2.0 \\
GRB 160408268 & 0.07 & 0.1  & 627 & 0.98 & 317 & -1.9 \\
GRB 160726065 & 0.08 & 6.5  & 276 & 0.85 & 247 & -0.2 \\
GRB 160804180 & 0.16 & 10.3 & 357 & 0.33 & 358 & -1.6 \\
GRB 160818198 & 0.6  & 8.3  & 407 & 0.57 & 13   & -0.8 \\
GRB 170709334 & 0.46 & 6.6  & 219 & 0.1  & 302 & 0.5  \\
GRB 170726794 & 0.28 & 9.1  & 337 & 0.59 & 418 & -0.3 \\
GRB 170802638 & 0.15 & 4.6  & 700 & 0.39 & 98  & -1.1 \\
GRB 180402406 & 0.03 & 5.9  & 120 & 0.99 & 249 & -1.0 \\
GRB 180511437 & 2.80  & 1.2  & 252 & 0.17 & 468 & 0.1  \\
GRB 181126413 & 0.72 & 4.1  & 277 & 0.3  & 220 & -1.6 \\
GRB 191221802 & 0.03 & 1.0  & 806 & 0.05 & 140 & -1.8 \\
	\hline
\end{tabular*}
	\footnotesize{$^a$ The samples and Tpre (i.e. the duration of precursor) are from \citealp{zhong2019precursors} and \citealp{wang2020stringent}. \\
	$^b$ Significance of the signal of precursor obtained from the Li-Ma formula \citep{li1983analysis}. \\
	$^c$ $BF_{\rm qpo}$ is the Bayes factor, the ${\rm In}BF_{\rm qpo}({\rm GP})$=3 corresponds approximately to a $p$-value of 0.001 \citep{hubner2022searching}.}\\
\end{table*}

\subsection{Data selection}

A precursor usually should satisfy the following requirements \citep{troja2010precursors,wang2020stringent}: (1) the first pulse in the GRB; (2) the peak count rate is lower than that of the main pulse; (3) the count rate during the waiting time period is consistent with the background level. We collect our samples for SGRBs and LGRBs with precursors observed by GBM from literature (\citealp{burlon2008precursors}; \citealp{hu2014internal}; \citealp{minaev2017precursors}; \citealp{zhong2019precursors}; \citealp{coppin2020identification}; \citealp{wang2020stringent}; \citealp{li2022temporal}). From 2008 to 2019, there are 24 samples of precursors for SGRBs (see Table \ref{SGRBs_table}) and 185 samples for LGRBs precursor.
% and filtered them by the above criteria

To improve the statistics, only those detectors of GBM for which the incident angle of the source is less than 90 degrees for SGRBs and 70 degrees for LGRBs are selected, and only the events in the energy range 8-900 keV are used in this analysis. It is worth noting that since most of GRBs are not precisely located, and especially for SGRBs with large localization errors, we use the GRB direction with the highest probability  (i.e. central value) as a reference for detector selection.

\subsection{Significance of the precursors}

Since the QPO search depends on the brightness of the signal (i.e. signal-to-noise ratio) (e.g. \citealp{lewin1988review}), we calculate the significance of the candidate precursor signal by Li-Ma formula \citep{li1983analysis}, which is defined as 
\begin{align}
& S = \sqrt{2} \left\{ n\log{\left[ \frac{\alpha + 1}{\alpha} 
\left(\frac{n}{n+b}\right)\right]} + b \log{ \left[ (\alpha +1) \frac{b}{n+b} \right] } \right\}^{1/2},
\label{eq:sign}
\end{align}
where $n$ is the counts of a certain time $t_{\rm on}$ within which a suspected source exists, $b$ is the measured background counts in a time
interval $t_{\rm off}$, and $\alpha$ is the ratio of the on-source time to the off-source time $t_{\rm on}/t_{\rm off}$.

On the other hand, dead time has an intricate effect on the periodogram \citep{huppenkothen2013quasi}. Fortunately, the dead time of a single normal event of GBM is only 2.6 $\mu s$ \citep{2009ApJ...702..791M}, and there are no extremely bright or saturated ones in our precursor sample, therefore the effect of dead time does not need to be considered in this work.

\begin{figure*}[http]
\centering
\begin{minipage}[t]{0.9\textwidth}
\centering
\includegraphics[width=\columnwidth]{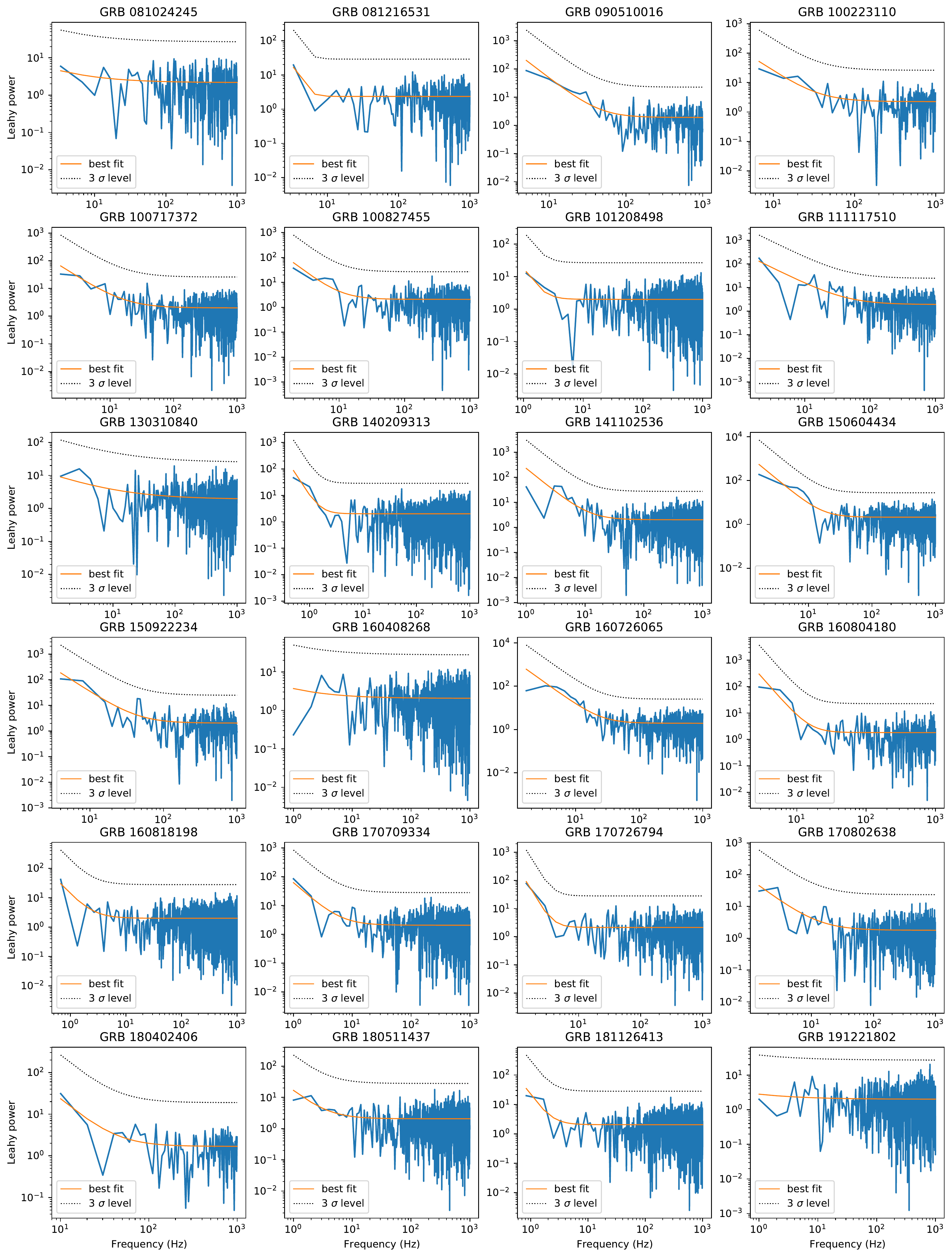}
\end{minipage}
\caption{PDS of precursors in SGRBs. The solid
orange lines show the best–fitting power law+constant models, and the black dotted lines represent the 3 $\sigma$ level. The number of frequencies searched is corrected.}\label{sgrb_fft}
\end{figure*}

\begin{figure*}
\centering
\begin{minipage}[t]{0.48\textwidth}
\centering
\includegraphics[width=\columnwidth]{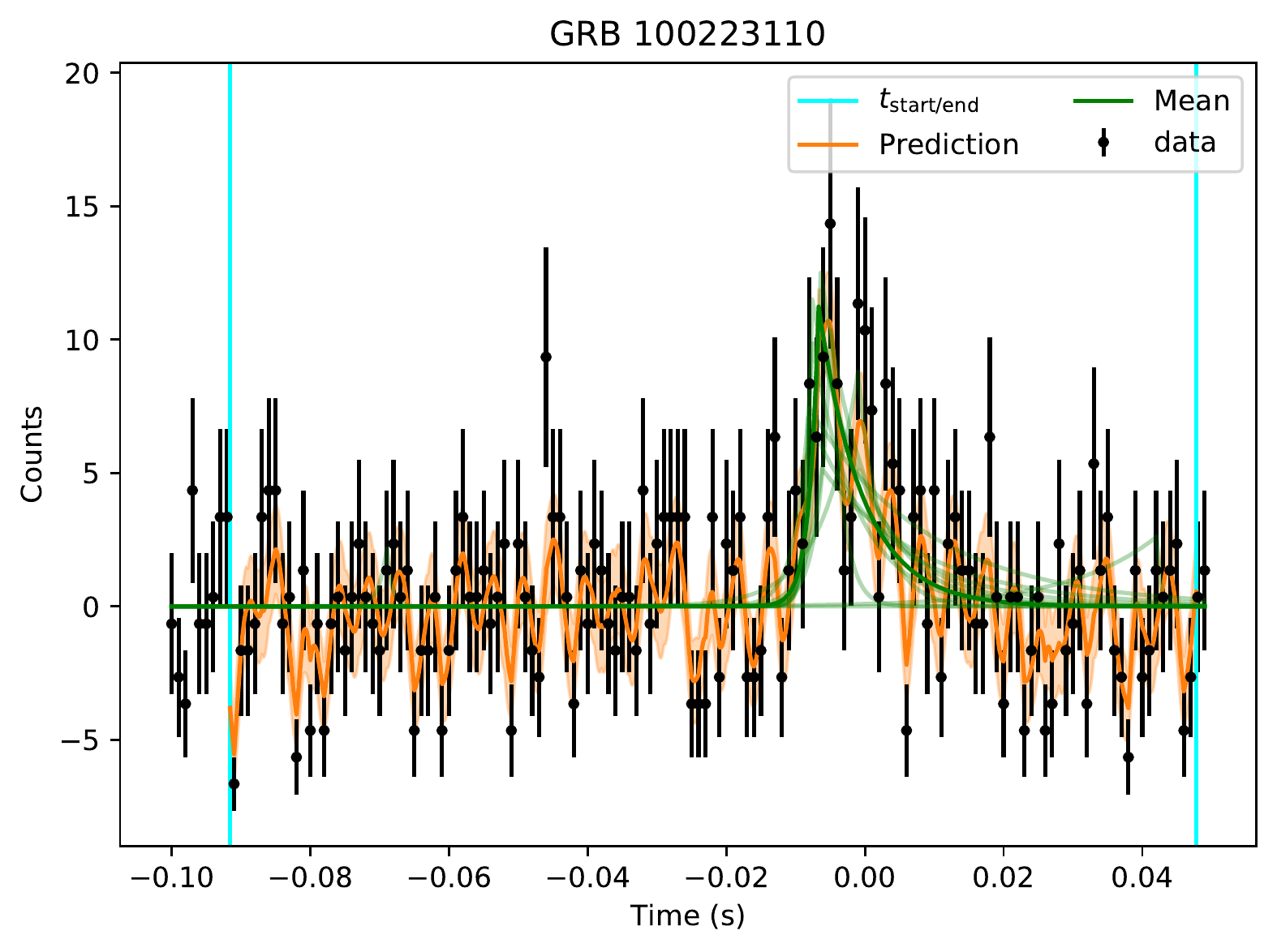}
\end{minipage}
\begin{minipage}[t]{0.48\textwidth}
\centering
\includegraphics[width=\columnwidth]{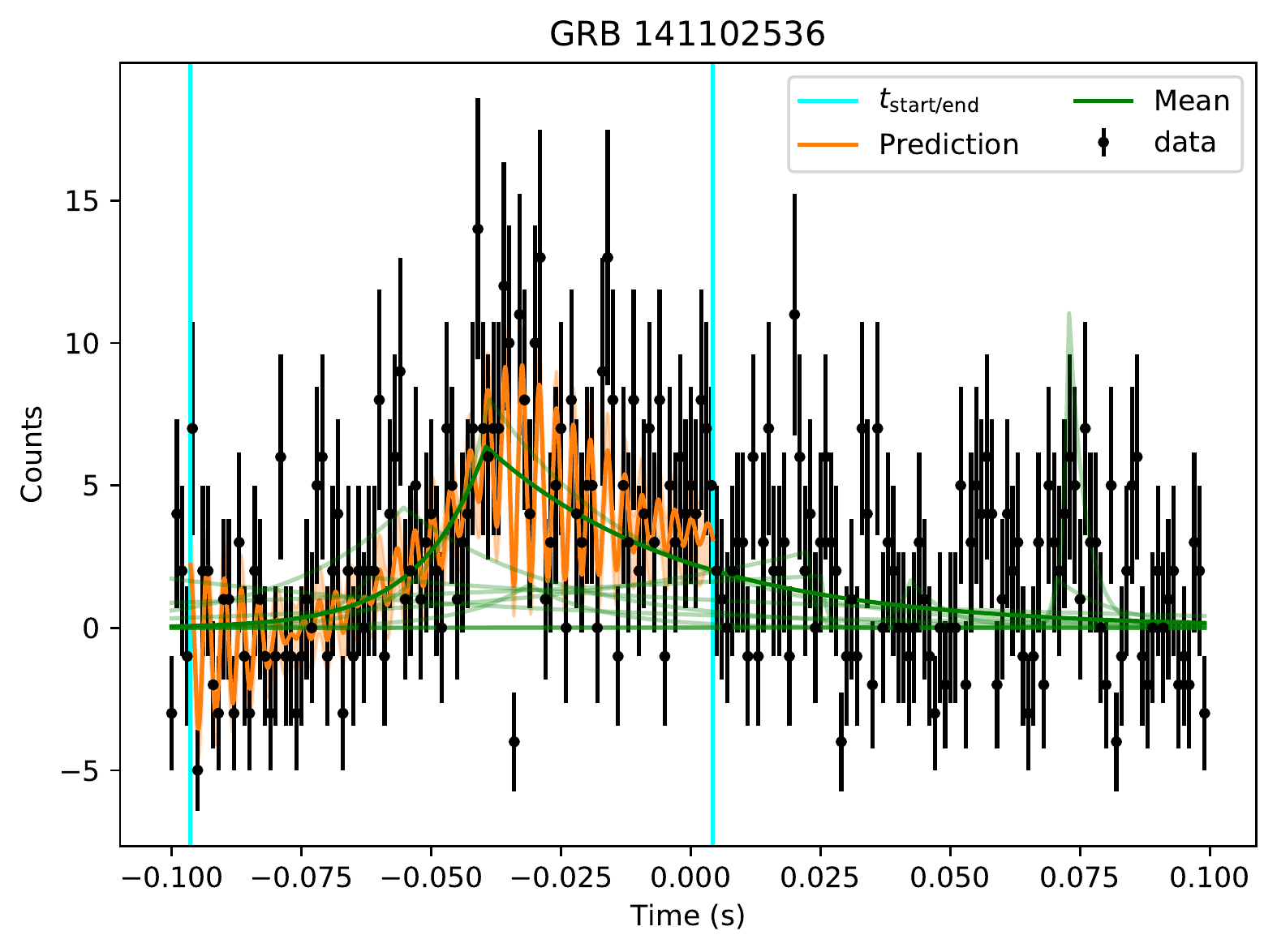}
\end{minipage}
\begin{minipage}[t]{0.48\textwidth}
\centering
\includegraphics[width=\columnwidth]{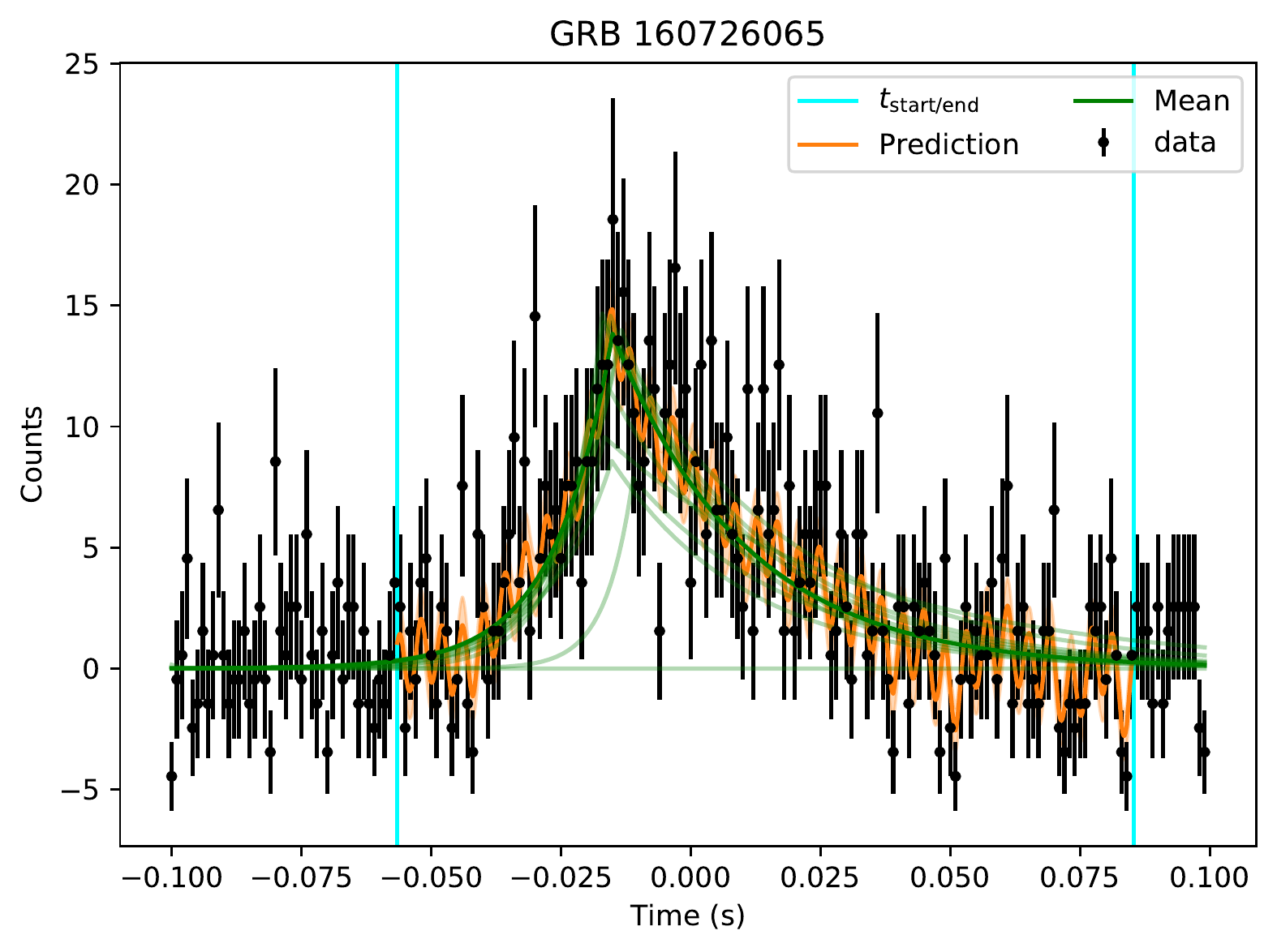}
\end{minipage}
\begin{minipage}[t]{0.48\textwidth}
\centering
\includegraphics[width=\columnwidth]{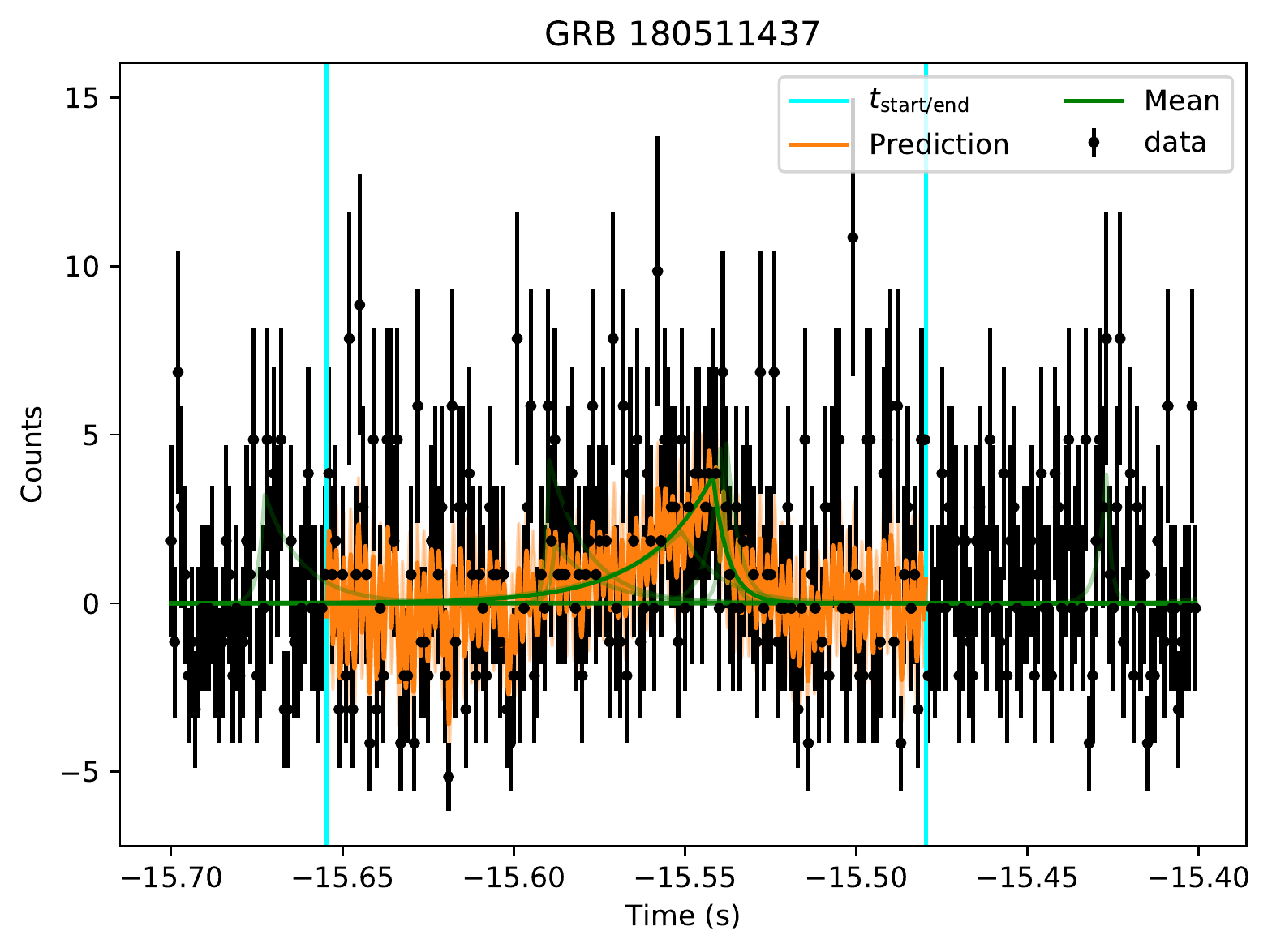}
\end{minipage}
\caption{The light curves of precursors in GRB 100223110, GRB 141102536, GRB 160726065 and GRB 180511437. The green lines are the mean functions (i.e. FRED) from
the maximum likelihood samples, the orange lines are the predictions based on the maximum likelihood samples, and the light blue vertical lines are the maximum likelihood start or end of the red noise and possible QPO. However, the ${\rm In}BF_{\rm qpo}$ of them are all less than 3, that is, there is not enough significance to confirm the existence of QPO. The time bin size of light curves are 1 ms.}\label{GP_sample}
\end{figure*}

\subsection{Search for periodic signals}

Most common methods for QPOs search are frequency-domain based methods such as periodograms or power spectra, which are based on assumption that bins in a periodogram are 
${\chi_{2}}^2$-distributed, thus a Whittle likelihood applies. It is only strictly true for infinitely long time series with homoscedastic stationary Gaussian data \citep{hubner2022searching}; however, for fast transients like GRBs, the statistical distributions at low frequencies are not stationary processes \citep{huppenkothen2013quasi,hubner2022pitfalls}. Recently, H{\"u}bner et al. proposed to search for QPOs in astrophysical transients based on Gaussian processes (GP) \citep{hubner2022searching}, which can investigate QPO directly in the time domain and take account of heteroscedasticity and non-stationarity in data. Here we use both methods,i.e., Power density spectrum (PDS) and Gaussian processes, to search for the QPO frequency from about 1 to 1000 Hz and to estimate the significance of the possible QPOs. 
Since the precursor of a SGRB usually has only one FRED-shaped pulse, both methods can be applied. However, the precursor of a LGRB usually has several pulses, thus we do not use the GP method for LGRBs.

\subsubsection{Power density spectrum method}
Within a Bayesian framework, we
model the periodogram as a combination of red noise at low frequencies and white noise at high frequencies with a power-law plus a constant  \citep{2010MNRAS.402..307V,guidorzi2016individual,huppenkothen2013quasi},
\begin{equation}
S_{\rm plc}(f) = \beta f^{-\alpha}+\gamma.
\label{eq:plc}
\end{equation}

When we have a parametric power spectral model with known
parameters $\theta$, $S_{j}(\theta)$, the ratio (or residuals) $R_{j}^{obj}=2I_{j}^{obs}/S_{j}(\theta)$ will follow a $\chi_{v}^2$ distribution with $v$ = 2 degrees of freedom \citep{vaughan2005simple}, and the highest outlier in the residuals is the candidate QPO.

To estimate the $p$-value (or significance), we simulate a large number of periodograms from the sample of Markov Chain Monte Carlo simulations (MCMC), compute the residuals and find the maximum outlier in the residuals for each periodogram. Finally, we calculate the $p$-value of the maximum outlier of the real burst in the distribution of maximum outliers from the set of simulations derived from the broadband noise model with no periodicity (see \citealp{huppenkothen2013quasi} for details). We perform some of the above calculations using $Stingray$ package \footnote{\href{https://docs.stingray.science/index.html}{https://docs.stingray.science/index.html}} \citep{huppenkothen2019stingray}. It is worth noting that the most conservative correction for the number of trials should be applied, that is, the $p$-values are corrected for the number of bursts searched and frequencies searched. However for the purpose of comparing the results of SGRBs and LGRBs, the significance reported in Table \ref{SGRBs_table} is only corrected for the number of frequencies searched but not for the number of bursts searched. To achieve a frequency search up to 1000 Hz, the time bin size of the light curves is 0.5 ms. 

\begin{figure}[http]
\centering
\includegraphics[width=\columnwidth]{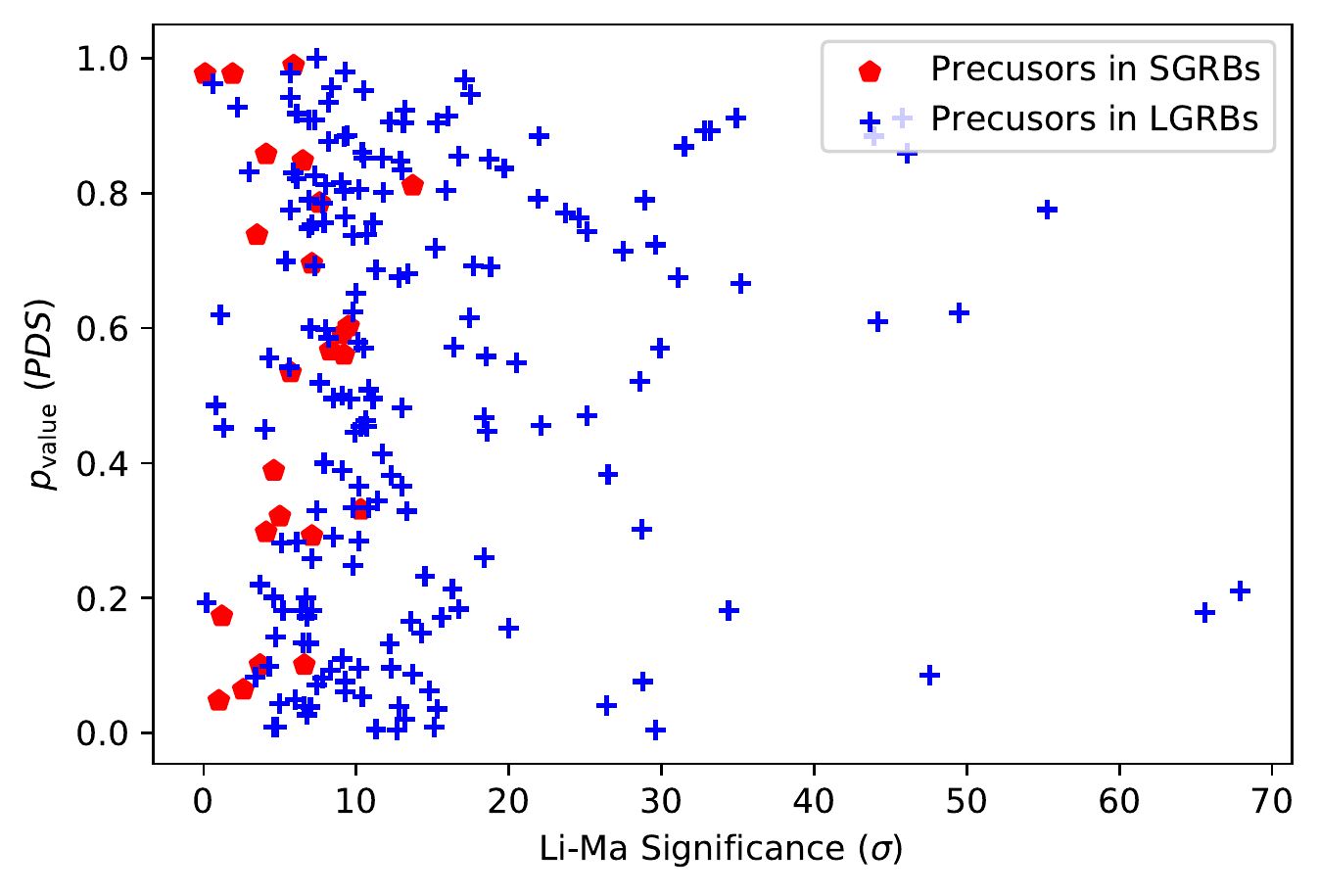}
\caption{The Li-Ma significance versus the $p_{\rm value}$(PDS) of potential QPO. The Pearson correlation coefficients for them are 0.17 and 0.08 for precursors of SGRBs and LGRBs, respectively. Thus no clues are found for the common presence of QPO in the precursors.}\label{sig_p}
\end{figure}

\begin{figure}[http]
\centering
\includegraphics[width=\columnwidth]{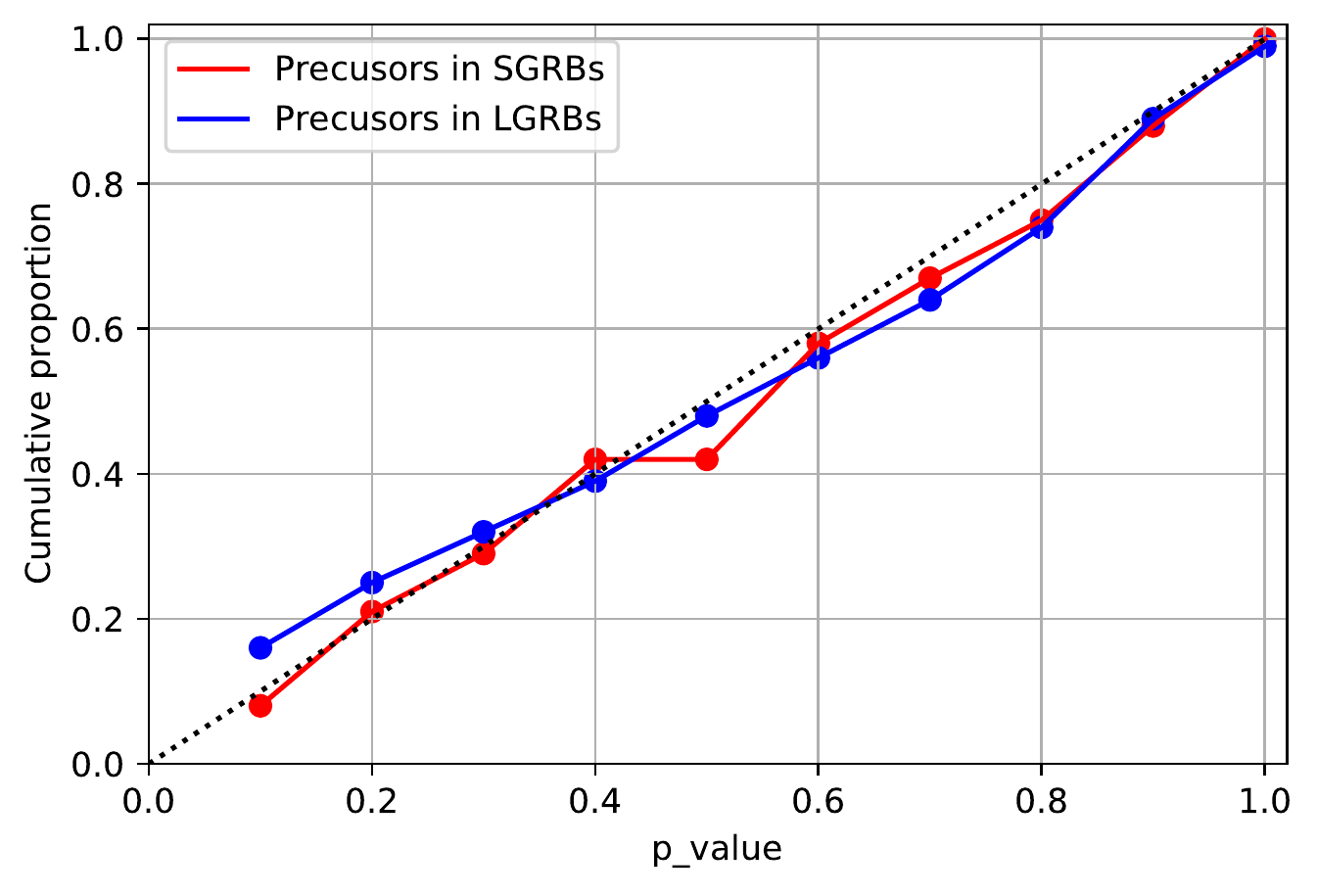}
\caption{$p$-value region versus the fraction of times that the $p$-value falls within $\leq p$-value for PDS method. The black dotted line represents the perfect one-to-one relation.}\label{valid}
\end{figure}

\subsubsection{Gaussian processes}
We adopt the procedure in \citealp{hubner2022searching}. It models QPOs as a stochastic process on top of a deterministic shape (i.e. mean function). For the deterministic shape, we find that a FRED can be better fitted to the precursors of SGRBs. For the stochastic process, we define the kernels describing a periodic oscillation and red noise as follows,
\begin{equation}
\begin{split}\label{equ:gs}
k_{\rm qpo}(\tau)= a_0 \cos(2\pi f\tau) \exp(-c_0\tau) ,\\
k_{\rm rn}(\tau) = a_1 \exp(-c_1\tau),
\end{split}
\end{equation}
where $a_0$ is the amplitude of the oscillation at frequency $f$, $c_0$ a description of the decay of the QPO with time constant $\tau$, $a_1$, and $c_1$ are the amplitude of red noise and the decay with time, respectively.

To assess the significance of the QPO, we perform model selection between QPOs and red noise (for more detailed information refer to \cite{hubner2022searching}), the Bayes factor $BF_{\rm qpo}$ is defined as
\begin{equation}
\begin{split}\label{equ:gs}
BF_{\rm qpo}=\frac{Z(d|k_{\rm qpo+rn},\mu)}{Z(d|k_{\rm rn},\mu)},
\end{split}
\end{equation}
where $Z(d|k_{\rm qpo+rn},\mu)$ (i.e. QPO and red noise) and $Z(d|k_{\rm rn},\mu$) (i.e. red noise) are the respective evidence in the different models, the $\mu$ is the parameter of the mean function and $d$ represents data. Finally, $Z$ is the evidence describing the overall probability that a given model for data,
\begin{equation}
\begin{split}\label{equ:gs}
Z(d| S) = \int \pi(\theta)L(d|\theta, S) d\theta, \\
L(d|\theta, M)= \frac{p(\theta|d, M)Z(d| M)}{\pi(\theta|M) },
\end{split}
\end{equation}
where $\theta$, $d$, $M$ and $p(\theta|d, M)$ are parameters, data, model and the posterior probability, respectively; $\pi(\theta|M)$ and $L(d|\theta, M)$ are the {\it prior} probability and likelihood, respectively. 

To improve the statistics (i.e. the statistical fluctuation in counts is large when the time binsize is too small), the time bin size of the light curves is set as 1 ms, which a frequency search up to 500 Hz. When $\ln BF_{\rm qpo}>3$, it is very unlikely that we have seen a false positive, which corresponds to a $p$-value of approximately 0.001 (see Fig. 5 in \citealp{hubner2022searching}). In this work, we refer to the publicly available code of GP  \footnote{\href{https://github.com/MoritzThomasHuebner/QPOEstimation}{https://github.com/MoritzThomasHuebner/QPOEstimation}} released by \citealp{hubner2022searching} and report the result in this work using non-stationary models.

\section{Result}

\subsection{Precursors in SGRBs}
We search for QPOs for the 24 SGRBs precursors using both the PDS and GP methods, and the results are shown in Table \ref{SGRBs_table}, note we only report the frequency with the highest deviation (i.e. minimum p-value) from the best fit model for that specific power density spectra. We found no candidate with significance above 3 $\sigma$ or $\ln BF_{\rm qpo}>3$. However, it is worth noting that we cannot exclude the possibility of the existence of QPOs in SGRBs precursors, as most of them are very weak (i.e. a few net counts). The Pearson correlation coefficient (see Fig.~\ref{sig_p}) for the Li-Ma significance and $p$-value of potential QPO is 0.17 ( smaller negative values represent higher significance of potential common QPO for precursors, that is, brighter precursors should have smaller p-values (or higher significance)), i.e., no clues are found for the common presence of QPOs in the precursors of SGRBs.

For the results obtained by the GP method, although all $\ln BF_{\rm qpo}$ are less than 3 (i.e. $< 3\sigma$), we note that some candidate QPO frequencies are consistent with the results obtained in the PDS (e.g. GRB 100223110). In Fig. \ref{GP_sample}, we show the light curves of four precursors for which the GP model fits the data relatively well; however, it is difficult to confirm the existence of QPOs due to the low statistics.

\subsection{Precursors in LGRBs}
In contrast to SGRBs, most precursors in LGRBs have sufficient net counts to test whether QPOs really exist. We search for QPOs for the 185 LGRB precursors using PDS, but also find no candidate with significance above 3 $\sigma$. The Pearson correlation coefficient for the Li-Ma significance and $p$-value of potential QPO is 0.08 (see Fig.~\ref{sig_p}), and thus no clues are found for the common presence of QPO in the precursors of LGRBs. 

In order to verify the robustness of the $p$-value estimation  in PDS method (in the Tabel.~\ref{LGRBs_table}), that is, whether the percentage of samples smaller than a certain $p$-value is this $p$-value, assuming that QPOs do not commonly exist. As shown by the blue solid line in Fig.~\ref{valid} (or the cumulative distribution of p-values obtained for best QPO candidate of each precursor), each quoted $p$-value contains samples with a percentage close to its nominal level, which means that the $p$-value estimation is robust. In addition, the $p$-value estimation of QPOs for SGRB precursors (in the Tabel.~\ref{SGRBs_table}) is also robust (see red solid line in Fig.~\ref{valid}).

\begin{figure}[http]
\centering
\includegraphics[width=\columnwidth]{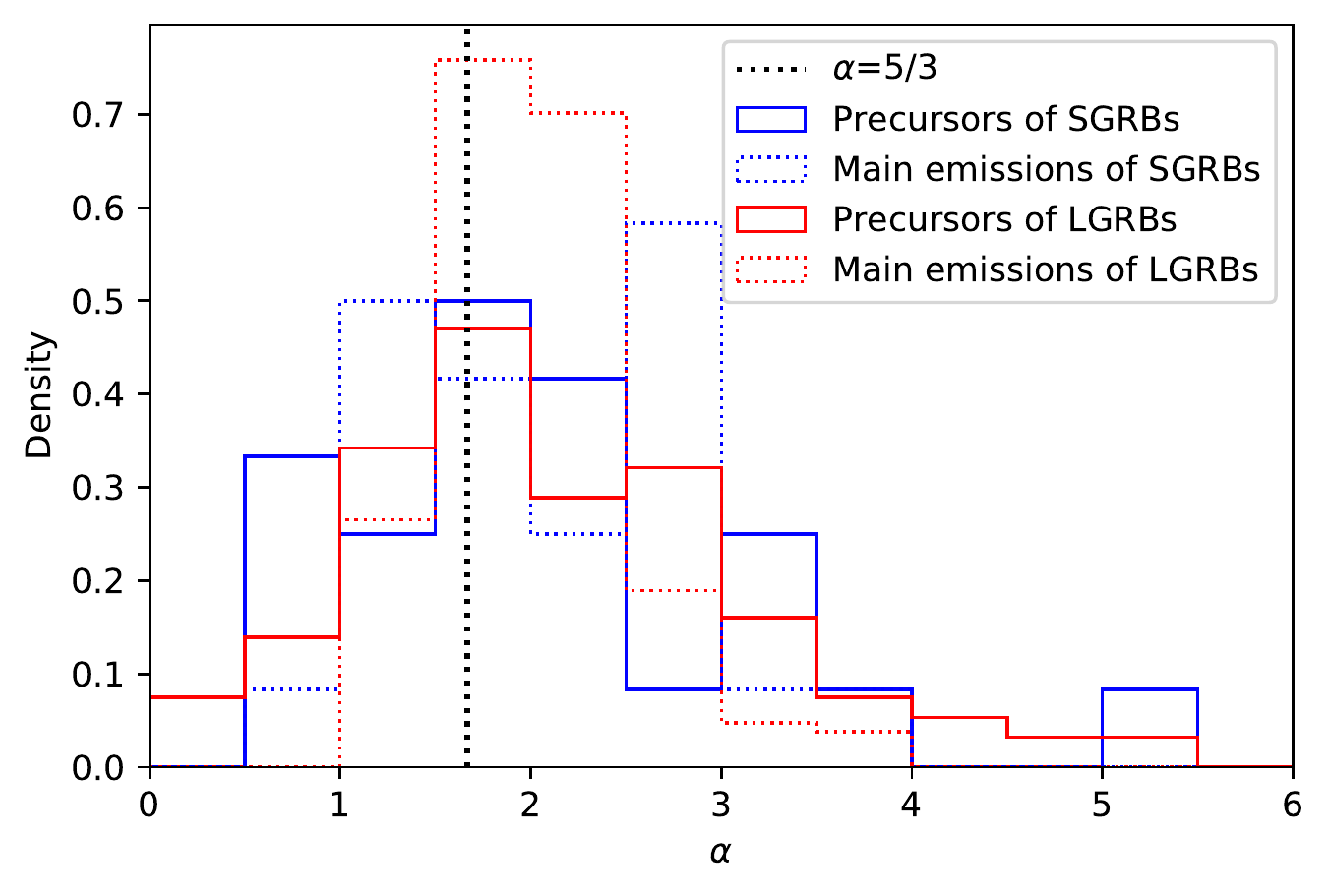}
\caption{Distribution of the PDS slope ($\alpha$) for precursors and main emissions of SGRBs and LGRBs.}\label{alphas}
\end{figure}

\section{Discussion and conclusion}
We do not find any convinced QPO signal (with significance $>3\sigma$) in the precursors of SGRBs and LGRBs observed by GBM from 2008 to 2019 based on the frequency-domain (i.e. PDS) and time-domain (i.e. GP) methods. We can rule out the possibility that QPOs commonly exist in LGRB precursors with their large sample and sufficient brightness. However, our results do not allow us to exclude the possibility of the common QPO existence in SGRB precursors due to their relatively small sample and weak signal. 

The Li-Ma significance levels of the individual precursors of SGRBs (e.g. GRB 160408268) are very low with the GBM data due to a few counts; these precursors were detected relatively significantly only by Swift/BAT. Besides, we classify LGRBs and SGRBs in this work only with their duration; however, since the ‘tip-of-iceberg’ effect \citep{lu2014amplitude} may make what would have been a long burst to be observed as a SGRB with a precursor. Further studies should classify GRBs by spectral lag and spectral shape, for example, SGRBs are commonly associated with harder spectra and spectral lags consistent with $\sim$0 s \citep{berger2014short}. In addition, we also find that in some of the light curves of individual samples taken from the literature the count rates in waiting times are not strictly the same as the background level, such as GRB 150922234 and GRB 160804180 (see \citealp{wang2020stringent}), that is, some SGRB precursors  may be also LGRBs. The above two factors means that the real sample of SGRBs may be even smaller. 

The lack of unambiguous evidence for QPO signal in the precursors of LGRBs does not conflict with the precursor model, in which the precursor is considered to be directly related to the central engine activities and shares the same physical origin with the prompt episode \citep{hu2014internal}. This is consistent with the fact that no confirmed candidates are found in the previous QPO searches in the main emissions of LGRBs \citep{dichiara2013search,huppenkothen2013quasi,guidorzi2016individual,dichiara2016correlation}. To test whether precursor QPOs are common and the significance is related to the brightness of a precursor, we also investigated the correlation between the precursor's Li-Ma significance and the $p$-value of a candidate QPO. The Pearson correlation coefficients for the precursors in SGRBs and LGRBs are 0.17 and 0.08, respectively, i.e., no clues are found for the common presence of QPOs in these precursors. Besides, we verified the robustness of the $p$-value estimation for the PDS method, that is, the percentage of samples smaller than a certain $p$-value is this $p$-value.

On the other hand, although some magnetar theories predict the possible existence of QPOs of tens of Hz in SGRB precursor (e.g. \citealp{sotani2007torsional,tews2017spectrum,zhang2022tidal}), the duration of the precursors of SGRBs is just about $\sim$0.1 s (see Table.\ref{SGRBs_table}), which means that only a few cycles of oscillations could possibly exist in a precursor, making the QPO search even more difficult. A definitive answer may come from sufficient statistics observed by more advanced detectors and higher detection signal to noise ratios, or a joined analysis using the light curves obtained by multiple GRB detectors, such as Fermi/GBM \citep{meegan2009fermi}, {\it Insight}-HXMT/HE \citep{zhang2020overview,liu2020high}, Swift/BAT \citep{barthelmy2005burst}, and GECAM (\citealp{li2021inflight,xiao2022ground}). Especially for detectors with similar energy response will be more advantageous, such as GECAM and GBM \citep{xiao2022energetic}.
Also can by combining with the information from theoretical models and the QPO evolution with time as a template in searching for QPOs \citep{dichiara2013search}. 
% Also can by combining with the information from gravitation wave radiation, for example, taking the frequency of the detected gravitational wave and its evolution with time as a template in searching for QPOs in the simultaneously detected light curve of the precursor. 

In addition, for the Gaussian process method, we used the light curves of 1 ms time bin size to achieve a frequency search of up to 500 Hz. However, we find that there is an optimal time binsize with simulated data containing QPOs, that is, the significance is lower when the time bin size is too small or too large. This may be similar to the phenomenon found in similar timing analysis \citep{ukwatta2010spectral,2021ApJ...920...43X,xiao2022robust}, that is, when the time binsize is too small, the statistical fluctuation in counts is large, and when the time bin is too large, the information in the light curves is lost. Therefore, we suggest that when possible QPO candidates exist (e.g. Significance obtained by PDS method is above 3 $\sigma$), an appropriate time binsize should be chosen.

Although the QPO search in this work has only yielded negative results, an interesting product of the detailed search is the continuum properties of the PDS for the precursors and main emissions, which are also studied in this work for the first time. The study can help to constrain the energy dissipation in GRBs as a stochastic process and also the dissipation region \citep{titarchuk2007power,vaughan2013random,guidorzi2016individual}.
Fig.\ref{alphas} shows the distribution of the power-law indices of PDS for the precursors and main emission of both SGRBs and LGRBs, and no significant difference is found, consistent with previous studies\citep{dichiara2013search,dichiara2016correlation}. The power-law indices are peaked at $\sim$ 5/3, which is similar to that of the hydrodynamics in the Kolmogorov spectrum of velocity fluctuations within a medium characterized by a fully developed turbulence \citep{beloborodov1998self}. 

In summary, we have performed systematic QPO search in the precursors of GRBs for the first time, which complements previous QPO studies. As well as reporting for the first time the distributions of the PDS of precursor and the relation with the main emission, which will contribute to further understanding of the physical mechanism of GRBs and their precursors (e.g. \citealp{narayan2009turbulent,2011ApJ...727L..41C}), as well as constraining the parameter space of the GRB internal shock model, such as the wind ejection properties determine the optical thickness of the wind in relativistic shells colliding with each other and dynamics (e.g. \citealp{1994ApJ...430L..93R,1999ApJ...522L.105P}).

\section*{Acknowledgments}
We thank the anonymous reviewer for a careful reading of our manuscript and insightful comments and suggestions. We acknowledge the public data from {\it Fermi}/GBM. The authors thank supports from 
the Strategic Priority Research Program on Space Science, the Chinese Academy of Sciences (Grant No.
XDA15010100, XDA15360100, XDA15360102, XDA15360300,%% GECAM SAS
XDA15052700) %% Xiong shaolin, GECAM data analysis
, the National Natural Science Foundation of China (Projects: 12061131007 and Grant No. 12173038).  %%Li Xinqiao, GECAM space environment 
This work was partially supported by International Partnership Program of Chinese Academy of Sciences (Grant No.113111KYSB20190020). S. Xiao is grateful to W. Xiao, G. Q. Wang and J. H. Li for their useful comments. 

\newpage

\bibliography{main}

\begin{thebibliography}{}
\expandafter\ifx\csname natexlab\endcsname\relax\def\natexlab#1{#1}\fi
\providecommand{\url}[1]{\href{#1}{#1}}
\providecommand{\dodoi}[1]{doi:~\href{http://doi.org/#1}{\nolinkurl{#1}}}
\providecommand{\doeprint}[1]{\href{http://ascl.net/#1}{\nolinkurl{http://ascl.net/#1}}}
\providecommand{\doarXiv}[1]{\href{https://arxiv.org/abs/#1}{\nolinkurl{https://arxiv.org/abs/#1}}}

\bibitem[{Abbott {et~al.}(2017)Abbott, Abbott, Abbott, Acernese, Ackley, Adams,
  Adams, Addesso, Adhikari, Adya, {et~al.}}]{abbott2017gravitational}
Abbott, B.~P., Abbott, R., Abbott, T., {et~al.} 2017, The Astrophysical Journal
  Letters, 848, L13

\bibitem[{Barthelmy {et~al.}(2005)Barthelmy, Barbier, Cummings, Fenimore,
  Gehrels, Hullinger, Krimm, Markwardt, Palmer, Parsons,
  {et~al.}}]{barthelmy2005burst}
Barthelmy, S.~D., Barbier, L.~M., Cummings, J.~R., {et~al.} 2005, Space Science
  Reviews, 120, 143

\bibitem[{Beloborodov {et~al.}(1998)Beloborodov, Stern, \&
  Svensson}]{beloborodov1998self}
Beloborodov, A.~M., Stern, B.~E., \& Svensson, R. 1998, The Astrophysical
  Journal, 508, L25

\bibitem[{Berger(2014)}]{berger2014short}
Berger, E. 2014, Annual review of Astronomy and Astrophysics, 52, 43

\bibitem[{Burlon {et~al.}(2008)Burlon, Ghirlanda, Ghisellini, Lazzati, Nava,
  Nardini, \& Celotti}]{burlon2008precursors}
Burlon, D., Ghirlanda, G., Ghisellini, G., {et~al.} 2008, The Astrophysical
  Journal, 685, L19

\bibitem[{{Carballido} \& {Lee}(2011)}]{2011ApJ...727L..41C}
{Carballido}, A., \& {Lee}, W.~H. 2011, \apjl, 727, L41,
  \dodoi{10.1088/2041-8205/727/2/L41}

\bibitem[{Cenko {et~al.}(2010)Cenko, Butler, Ofek, Perley, Morgan, Frail,
  Gorosabel, Bloom, Castro-Tirado, Cepa, {et~al.}}]{cenko2010unveiling}
Cenko, S., Butler, N., Ofek, E., {et~al.} 2010, The Astronomical Journal, 140,
  224

\bibitem[{Coppin {et~al.}(2020)Coppin, de~Vries, \& van
  Eijndhoven}]{coppin2020identification}
Coppin, P., de~Vries, K.~D., \& van Eijndhoven, N. 2020, Physical Review D,
  102, 103014

\bibitem[{Dichiara {et~al.}(2016)Dichiara, Guidorzi, Amati, Frontera, \&
  Margutti}]{dichiara2016correlation}
Dichiara, S., Guidorzi, C., Amati, L., Frontera, F., \& Margutti, R. 2016,
  Astronomy \& Astrophysics, 589, A97

\bibitem[{Dichiara {et~al.}(2013)Dichiara, Guidorzi, Frontera, \&
  Amati}]{dichiara2013search}
Dichiara, S., Guidorzi, C., Frontera, F., \& Amati, L. 2013, The Astrophysical
  Journal, 777, 132

\bibitem[{Gao {et~al.}(2022)Gao, Lei, \& Zhu}]{gao2022grb}
Gao, H., Lei, W.-H., \& Zhu, Z.-P. 2022, arXiv preprint arXiv:2205.05031

\bibitem[{Guidorzi {et~al.}(2016)Guidorzi, Dichiara, \&
  Amati}]{guidorzi2016individual}
Guidorzi, C., Dichiara, S., \& Amati, L. 2016, Astronomy \& Astrophysics, 589,
  A98

\bibitem[{Hansen \& Lyutikov(2001)}]{hansen2001radio}
Hansen, B.~M., \& Lyutikov, M. 2001, Monthly Notices of the Royal Astronomical
  Society, 322, 695

\bibitem[{Hu {et~al.}(2014)Hu, Liang, Xi, Peng, Lu, L{\"u}, \&
  Zhang}]{hu2014internal}
Hu, Y.-D., Liang, E.-W., Xi, S.-Q., {et~al.} 2014, The Astrophysical Journal,
  789, 145

\bibitem[{H{\"u}bner {et~al.}(2022{\natexlab{a}})H{\"u}bner, Huppenkothen,
  Lasky, \& Inglis}]{hubner2022pitfalls}
H{\"u}bner, M., Huppenkothen, D., Lasky, P.~D., \& Inglis, A.~R.
  2022{\natexlab{a}}, The Astrophysical Journal Supplement Series, 259, 32

\bibitem[{H{\"u}bner {et~al.}(2022{\natexlab{b}})H{\"u}bner, Huppenkothen,
  Lasky, Inglis, Ick, \& Hogg}]{hubner2022searching}
H{\"u}bner, M., Huppenkothen, D., Lasky, P.~D., {et~al.} 2022{\natexlab{b}},
  arXiv preprint arXiv:2205.12716

\bibitem[{Huppenkothen {et~al.}(2013)Huppenkothen, Watts, Uttley, Van~der
  Horst, Van~der Klis, Kouveliotou, G{\"o}{\u{g}}{\"u}{\c{s}}, Granot, Vaughan,
  \& Finger}]{huppenkothen2013quasi}
Huppenkothen, D., Watts, A.~L., Uttley, P., {et~al.} 2013, The Astrophysical
  Journal, 768, 87

\bibitem[{Huppenkothen {et~al.}(2019)Huppenkothen, Bachetti, Stevens, Migliari,
  Balm, Hammad, Khan, Mishra, Rashid, Sharma,
  {et~al.}}]{huppenkothen2019stingray}
Huppenkothen, D., Bachetti, M., Stevens, A.~L., {et~al.} 2019, The
  Astrophysical Journal, 881, 39

\bibitem[{Kruger {et~al.}(2002)Kruger, Loredo, \& Wasserman}]{kruger2002search}
Kruger, A.~T., Loredo, T.~J., \& Wasserman, I. 2002, The Astrophysical Journal,
  576, 932

\bibitem[{Lewin {et~al.}(1988)Lewin, Van~Paradijs, \& Van~der
  Klis}]{lewin1988review}
Lewin, W.~H., Van~Paradijs, J., \& Van~der Klis, M. 1988, Space Science
  Reviews, 46, 273

\bibitem[{Li \& Mao(2022)}]{li2022temporal}
Li, L., \& Mao, J. 2022, The Astrophysical Journal, 928, 152

\bibitem[{Li \& Paczy{\'n}ski(1998)}]{li1998transient}
Li, L.-X., \& Paczy{\'n}ski, B. 1998, The Astrophysical Journal, 507, L59

\bibitem[{Li \& Ma(1983)}]{li1983analysis}
Li, T.-P., \& Ma, Y.-Q. 1983, The Astrophysical Journal, 272, 317

\bibitem[{Li {et~al.}(2021)Li, Wen, Xiong, Gong, Zhang, An, Xu, Liu, Cai,
  Chang, {et~al.}}]{li2021inflight}
Li, X., Wen, X., Xiong, S., {et~al.} 2021, arXiv preprint arXiv:2112.04772

\bibitem[{Liu {et~al.}(2020)Liu, Zhang, Li, Lu, Chang, Li, Zhang, Jin, Yu,
  Zhang, {et~al.}}]{liu2020high}
Liu, C., Zhang, Y., Li, X., {et~al.} 2020, SCIENCE CHINA Physics, Mechanics \&
  Astronomy, 63, 1

\bibitem[{L{\"u} {et~al.}(2014)L{\"u}, Zhang, Liang, Zhang, \&
  Sakamoto}]{lu2014amplitude}
L{\"u}, H.-J., Zhang, B., Liang, E.-W., Zhang, B.-B., \& Sakamoto, T. 2014,
  Monthly Notices of the Royal Astronomical Society, 442, 1922

\bibitem[{Mangan {et~al.}(2021)Mangan, Dunwoody, Meegan, Team,
  {et~al.}}]{mangan2021grb}
Mangan, J., Dunwoody, R., Meegan, C., Team, F.~G., {et~al.} 2021, GRB
  Coordinates Network, 31210, 1

\bibitem[{{Markwardt} {et~al.}(2009){Markwardt}, {Gavriil}, {Palmer},
  {Baumgartner}, \& {Barthelmy}}]{2009GCN..9645....1M}
{Markwardt}, C.~B., {Gavriil}, F.~P., {Palmer}, D.~M., {Baumgartner}, W.~H., \&
  {Barthelmy}, S.~D. 2009, GRB Coordinates Network, 9645, 1

\bibitem[{{Meegan} {et~al.}(2009){Meegan}, {Lichti}, {Bhat}, {Bissaldi},
  {Briggs}, {Connaughton}, {Diehl}, {Fishman}, {Greiner}, {Hoover}, {van der
  Horst}, {von Kienlin}, {Kippen}, {Kouveliotou}, {McBreen}, {Paciesas},
  {Preece}, {Steinle}, {Wallace}, {Wilson}, \&
  {Wilson-Hodge}}]{2009ApJ...702..791M}
{Meegan}, C., {Lichti}, G., {Bhat}, P.~N., {et~al.} 2009, \apj, 702, 791,
  \dodoi{10.1088/0004-637X/702/1/791}

\bibitem[{Meegan {et~al.}(2009)Meegan, Lichti, Bhat, Bissaldi, Briggs,
  Connaughton, Diehl, Fishman, Greiner, Hoover, {et~al.}}]{meegan2009fermi}
Meegan, C., Lichti, G., Bhat, P., {et~al.} 2009, The Astrophysical Journal,
  702, 791

\bibitem[{Minaev \& Pozanenko(2017)}]{minaev2017precursors}
Minaev, P.~Y., \& Pozanenko, A. 2017, Astronomy Letters, 43, 1

\bibitem[{Narayan \& Kumar(2009)}]{narayan2009turbulent}
Narayan, R., \& Kumar, P. 2009, Monthly Notices of the Royal Astronomical
  Society: Letters, 394, L117

\bibitem[{Palenzuela {et~al.}(2013)Palenzuela, Lehner, Ponce, Liebling,
  Anderson, Neilsen, \& Motl}]{palenzuela2013electromagnetic}
Palenzuela, C., Lehner, L., Ponce, M., {et~al.} 2013, Physical review letters,
  111, 061105

\bibitem[{{Panaitescu} {et~al.}(1999){Panaitescu}, {Spada}, \&
  {M{\'e}sz{\'a}ros}}]{1999ApJ...522L.105P}
{Panaitescu}, A., {Spada}, M., \& {M{\'e}sz{\'a}ros}, P. 1999, \apjl, 522,
  L105, \dodoi{10.1086/312230}

\bibitem[{Rastinejad {et~al.}(2022)Rastinejad, Gompertz, Levan, Fong, Nicholl,
  Lamb, Malesani, Nugent, Oates, Tanvir, {et~al.}}]{rastinejad2022kilonova}
Rastinejad, J., Gompertz, B., Levan, A., {et~al.} 2022, arXiv preprint
  arXiv:2204.10864

\bibitem[{{Rees} \& {Meszaros}(1994)}]{1994ApJ...430L..93R}
{Rees}, M.~J., \& {Meszaros}, P. 1994, \apjl, 430, L93, \dodoi{10.1086/187446}

\bibitem[{Sarin \& Lasky(2021)}]{sarin2021evolution}
Sarin, N., \& Lasky, P.~D. 2021, General Relativity and Gravitation, 53, 1

\bibitem[{Sotani {et~al.}(2007)Sotani, Kokkotas, \&
  Stergioulas}]{sotani2007torsional}
Sotani, H., Kokkotas, K., \& Stergioulas, N. 2007, Monthly Notices of the Royal
  Astronomical Society, 375, 261

\bibitem[{Sridhar {et~al.}(2021)Sridhar, Zrake, Metzger, Sironi, \&
  Giannios}]{sridhar2021shock}
Sridhar, N., Zrake, J., Metzger, B.~D., Sironi, L., \& Giannios, D. 2021,
  Monthly Notices of the Royal Astronomical Society, 501, 3184

\bibitem[{Stone {et~al.}(2013)Stone, Loeb, \& Berger}]{stone2013pulsations}
Stone, N., Loeb, A., \& Berger, E. 2013, Physical Review D, 87, 084053

\bibitem[{Suvorov {et~al.}(2022)Suvorov, Kuan, \& Kokkotas}]{suvorov2022quasi}
Suvorov, A.~G., Kuan, H.-J., \& Kokkotas, K.~D. 2022, arXiv preprint
  arXiv:2205.11112

\bibitem[{{Suvorov} {et~al.}(2022){Suvorov}, {Kuan}, \&
  {Kokkotas}}]{2022arXiv220511112S}
{Suvorov}, A.~G., {Kuan}, H.-J., \& {Kokkotas}, K.~D. 2022, arXiv e-prints,
  arXiv:2205.11112.
\newblock \doarXiv{2205.11112}

\bibitem[{Tarnopolski \& Marchenko(2021)}]{tarnopolski2021comprehensive}
Tarnopolski, M., \& Marchenko, V. 2021, The Astrophysical Journal, 911, 20

\bibitem[{Tews(2017)}]{tews2017spectrum}
Tews, I. 2017, Physical Review C, 95, 015803

\bibitem[{Titarchuk {et~al.}(2007)Titarchuk, Shaposhnikov, \&
  Arefiev}]{titarchuk2007power}
Titarchuk, L., Shaposhnikov, N., \& Arefiev, V. 2007, The Astrophysical
  Journal, 660, 556

\bibitem[{Troja {et~al.}(2010)Troja, Rosswog, \& Gehrels}]{troja2010precursors}
Troja, E., Rosswog, S., \& Gehrels, N. 2010, The Astrophysical Journal, 723,
  1711

\bibitem[{Troja {et~al.}(2022)Troja, Fryer, O'Connor, Ryan, Dichiara, Kumar,
  Ito, Gupta, Wollaeger, Norris, {et~al.}}]{troja2022long}
Troja, E., Fryer, C., O'Connor, B., {et~al.} 2022, arXiv preprint
  arXiv:2209.03363

\bibitem[{{Tsang} {et~al.}(2012){Tsang}, {Read}, {Hinderer}, {Piro}, \&
  {Bondarescu}}]{tsang2012resonant}
{Tsang}, D., {Read}, J.~S., {Hinderer}, T., {Piro}, A.~L., \& {Bondarescu}, R.
  2012, \prl, 108, 011102, \dodoi{10.1103/PhysRevLett.108.011102}

\bibitem[{Ukwatta {et~al.}(2010)Ukwatta, Stamatikos, Dhuga, Sakamoto,
  Barthelmy, Eskandarian, Gehrels, Maximon, Norris, \&
  Parke}]{ukwatta2010spectral}
Ukwatta, T., Stamatikos, M., Dhuga, K., {et~al.} 2010, The Astrophysical
  Journal, 711, 1073

\bibitem[{Vaughan(2005)}]{vaughan2005simple}
Vaughan, S. 2005, Astronomy \& Astrophysics, 431, 391

\bibitem[{{Vaughan}(2010)}]{2010MNRAS.402..307V}
{Vaughan}, S. 2010, \mnras, 402, 307, \dodoi{10.1111/j.1365-2966.2009.15868.x}

\bibitem[{Vaughan(2013)}]{vaughan2013random}
Vaughan, S. 2013, Philosophical Transactions of the Royal Society A:
  Mathematical, Physical and Engineering Sciences, 371, 20110549

\bibitem[{Wang {et~al.}(2018)Wang, Peng, Wu, \& Dai}]{wang2018pre}
Wang, J.-S., Peng, F.-K., Wu, K., \& Dai, Z.-G. 2018, The Astrophysical
  Journal, 868, 19

\bibitem[{Wang {et~al.}(2020)Wang, Peng, Zou, Zhang, \&
  Zhang}]{wang2020stringent}
Wang, J.-S., Peng, Z.-K., Zou, J.-H., Zhang, B.-B., \& Zhang, B. 2020, The
  Astrophysical Journal Letters, 902, L42

\bibitem[{Woosley \& Bloom(2006)}]{woosley2006supernova}
Woosley, S., \& Bloom, J. 2006, Annu. Rev. Astron. Astrophys., 44, 507

\bibitem[{Wu {et~al.}(2017)Wu, Tamborra, Just, \& Janka}]{wu2017imprints}
Wu, M.-R., Tamborra, I., Just, O., \& Janka, H.-T. 2017, Physical Review D, 96,
  123015

\bibitem[{{Xiao} {et~al.}(2021){Xiao}, {Xiong}, {Zhang}, {Song}, {Lu}, {Huang},
  {Cai}, {Yi}, {Song}, {Chen}, {Ge}, {Liu}, {Li}, {Li}, \&
  {Zhao}}]{2021ApJ...920...43X}
{Xiao}, S., {Xiong}, S.~L., {Zhang}, S.~N., {et~al.} 2021, \apj, 920, 43,
  \dodoi{10.3847/1538-4357/ac1420}

\bibitem[{Xiao {et~al.}(2022{\natexlab{a}})Xiao, Zhang, Zhu, Xiong, Gao, Xu,
  Zhang, Peng, Li, Zhang, {et~al.}}]{xiao2022quasi}
Xiao, S., Zhang, Y.-Q., Zhu, Z.-P., {et~al.} 2022{\natexlab{a}}, arXiv preprint
  arXiv:2205.02186

\bibitem[{Xiao {et~al.}(2022{\natexlab{b}})Xiao, Xiong, Wang, Zhang, Gao,
  Zhang, Cai, Yi, Zhao, Tuo, {et~al.}}]{xiao2022robust}
Xiao, S., Xiong, S.-L., Wang, Y., {et~al.} 2022{\natexlab{b}}, The
  Astrophysical Journal Letters, 924, L29

\bibitem[{Xiao {et~al.}(2022{\natexlab{c}})Xiao, Liu, Peng, An, Xiong, Tuo,
  Gong, Zhang, Zhang, Zheng, {et~al.}}]{xiao2022ground}
Xiao, S., Liu, Y., Peng, W., {et~al.} 2022{\natexlab{c}}, Monthly Notices of
  the Royal Astronomical Society, 511, 964

\bibitem[{Xiao {et~al.}(2022{\natexlab{d}})Xiao, Xiong, Cai, Song, Zheng, Peng,
  Wang, Qiao, Guo, Wang, {et~al.}}]{xiao2022energetic}
Xiao, S., Xiong, S.-L., Cai, C., {et~al.} 2022{\natexlab{d}}, Monthly Notices
  of the Royal Astronomical Society

\bibitem[{Yang {et~al.}(2022)Yang, Zhang, Ai, Liu, Wang, Li, L{\"u}, \&
  Zhang}]{yang2022peculiarly}
Yang, J., Zhang, B.-B., Ai, S., {et~al.} 2022, arXiv preprint arXiv:2204.12771

\bibitem[{Zhang {et~al.}(2020)Zhang, Li, Lu, Song, Xu, Liu, Chen, Cao, Bu,
  Chang, {et~al.}}]{zhang2020overview}
Zhang, S.-N., Li, T., Lu, F., {et~al.} 2020, SCIENCE CHINA Physics, Mechanics
  \& Astronomy, 63, 1

\bibitem[{Zhang {et~al.}(2022)Zhang, Yi, Zhang, Xiong, \&
  Xiao}]{zhang2022tidal}
Zhang, Z., Yi, S.-X., Zhang, S.-N., Xiong, S.-L., \& Xiao, S. 2022, arXiv
  preprint arXiv:2207.12324

\bibitem[{Zhilyaev \& Dubinovska(2009)}]{zhilyaev2009detection}
Zhilyaev, B., \& Dubinovska, D. 2009, Astronomische Nachrichten: Astronomical
  Notes, 330, 404

\bibitem[{Zhong {et~al.}(2019)Zhong, Dai, Cheng, Lan, \&
  Zhang}]{zhong2019precursors}
Zhong, S.-Q., Dai, Z.-G., Cheng, J.-G., Lan, L., \& Zhang, H.-M. 2019, The
  Astrophysical Journal, 884, 25

\end{thebibliography}

\appendix
\setcounter{table}{0}  
\renewcommand\thetable{\Alph{section}S\arabic{table}}

\begin{longtable*}{ccccc}
\endfirsthead
\hline
\endhead
\hline
\endfoot
\hline
\endlastfoot\\
\caption{\centering QPOs search for precursors in LGRBs using Power density spectrum. The Tpre is from \citealp{coppin2020identification}. }\label{LGRBs_table}\\
\hline
Name& Tpre (s) & Li-Ma Significance ($\sigma$)& $f{\rm _{PDS}}$ (Hz) &$p_{\rm value}({\rm PDS})$   \\

\hline
GRB 080723557 & 28.3  & 45.8 & 603 & 0.911 \\
GRB 080807993 & 1.0   & 34.9 & 534 & 0.911 \\
GRB 080816503 & 1.8   & 14.3 & 481 & 0.148 \\
GRB 080818579 & 5.6   & 9.2  & 981 & 0.803 \\
GRB 080830368 & 5.1   & 8.2  & 63  & 0.934 \\
GRB 081003644 & 4.3   & 8.2  & 59  & 0.586 \\
GRB 081121858 & 7.9   & 6.9  & 91  & 0.790 \\
GRB 090101758 & 6.1   & 7.9  & 776 & 0.756 \\
GRB 090113778 & 0.1   & 0.8  & 780 & 0.486 \\
GRB 090117335 & 1.3   & 9.3  & 806 & 0.077 \\
GRB 090131090 & 12.4  & 47.6 & 688 & 0.086 \\
GRB 090309767 & 6.1   & 10.8 & 337 & 0.334 \\
GRB 090419997 & 23.3  & 10.7 & 890 & 0.454 \\
GRB 090425377 & 2.7   & 10.8 & 862 & 0.509 \\
GRB 090502777 & 3.1   & 5.2  & 180 & 0.181 \\
GRB 090610723 & 6.7   & 9.1  & 217 & 0.500 \\
GRB 090618353 & 28.9  & 31.5 & 966 & 0.869 \\
GRB 090810659 & 43.3  & 4.7  & 375 & 0.142 \\
GRB 090811696 & 1.6   & 11.1 & 807 & 0.756 \\
GRB 090814950 & 18.6  & 7.6  & 758 & 0.519 \\
GRB 090820509 & 4.1   & 12.3 & 828 & 0.097 \\
GRB 090907017 & 1.7   & 9.3  & 37  & 0.061 \\
GRB 090929190 & 0.1   & 13.0 & 378 & 0.834 \\
GRB 091109895 & 2.8   & 16.4 & 240 & 0.572 \\
GRB 100116897 & 6.3   & 13.3 & 630 & 0.329 \\
GRB 100130729 & 23.2  & 6.6  & 773 & 0.040 \\
GRB 100204566 & 15.7  & 3.0  & 729 & 0.831 \\
GRB 100323542 & 9.1   & 5.7  & 594 & 0.978 \\
GRB 100517154 & 1.4   & 26.5 & 32  & 0.383 \\
GRB 100619015 & 9.9   & 13.6 & 812 & 0.166 \\
GRB 100625891 & 4.0   & 7.0  & 643 & 0.039 \\
GRB 100709602 & 16.3  & 10.2 & 721 & 0.806 \\
GRB 100718160 & 6.8   & 12.2 & 484 & 0.132 \\
GRB 100923844 & 4.0   & 4.6  & 525 & 0.009 \\
GRB 101224578 & 10.7  & 19.7 & 327 & 0.837 \\
GRB 101227536 & 3.6   & 29.6 & 360 & 0.004 \\
GRB 110102788 & 25.3  & 32.8 & 174 & 0.892 \\
GRB 110227229 & 21.1  & 4.8  & 629 & 0.009 \\
GRB 110428338 & 13.4  & 16.3 & 708 & 0.213 \\
GRB 110528624 & 13.8  & 1.3  & 762 & 0.452 \\
GRB 110725236 & 7.6   & 7.3  & 660 & 0.826 \\
GRB 110729142 & 51.6  & 12.7 & 300 & 0.005 \\
GRB 110825102 & 0.8   & 28.7 & 783 & 0.302 \\
GRB 110903111 & 22.1  & 29.9 & 815 & 0.570 \\
GRB 110904124 & 7.7   & 12.9 & 963 & 0.847 \\
GRB 110909116 & 1.7   & 16.7 & 423 & 0.854 \\
GRB 111010709 & 31.0  & 17.4 & 552 & 0.615 \\
GRB 111015427 & 17.1  & 3.7  & 359 & 0.220 \\
GRB 111230683 & 4.6   & 6.1  & 169 & 0.821 \\
GRB 111230819 & 4.2   & 8.0  & 834 & 0.598 \\
GRB 120308588 & 3.1   & 11.7 & 658 & 0.414 \\
GRB 120319983 & 5.6   & 7.4  & 272 & 0.071 \\
GRB 120412920 & 5.5   & 25.1 & 598 & 0.743 \\
GRB 120504945 & 0.8   & 7.9  & 75  & 0.399 \\
GRB 120530121 & 8.0   & 22.0 & 506 & 0.884 \\
GRB 120611108 & 6.6   & 5.9  & 460 & 0.830 \\
GRB 120710100 & 4.9   & 9.3  & 76  & 0.764 \\
GRB 120711115 & 4.8   & 9.8  & 88  & 0.249 \\
GRB 120716712 & 5.4   & 18.7 & 726 & 0.850 \\
GRB 120819048 & 1.6   & 6.9  & 308 & 0.748 \\
GRB 121005340 & 38.8  & 5.7  & 263 & 0.774 \\
GRB 121029350 & 8.8   & 13.0 & 177 & 0.482 \\
GRB 121031949 & 38.5  & 9.8  & 814 & 0.334 \\
GRB 121113544 & 31.7  & 18.6 & 461 & 0.447 \\
GRB 121125356 & 20.3  & 9.1  & 740 & 0.110 \\
GRB 121217313 & 65.8  & 4.0  & 256 & 0.450 \\
GRB 130104721 & 3.9   & 17.7 & 167 & 0.692 \\
GRB 130106995 & 17.6  & 16.0 & 841 & 0.914 \\
GRB 130208684 & 5.1   & 10.0 & 896 & 0.651 \\
GRB 130209961 & 4.6   & 33.2 & 678 & 0.892 \\
GRB 130219775 & 20.3  & 8.3  & 545 & 0.093 \\
GRB 130310840 & 1.2   & 9.9  & 747 & 0.446 \\
GRB 130318456 & 6.9   & 12.8 & 881 & 0.675 \\
GRB 130320560 & 42.1  & 9.2  & 453 & 0.884 \\
GRB 130404840 & 8.4   & 34.4 & 662 & 0.182 \\
GRB 130418844 & 16.5  & 10.2 & 624 & 0.284 \\
GRB 130623130 & 1.8   & 8.2  & 280 & 0.877 \\
GRB 130720582 & 115.4 & 23.7 & 341 & 0.771 \\
GRB 130813791 & 1.7   & 11.1 & 939 & 0.496 \\
GRB 130815660 & 6.9   & 20.0 & 551 & 0.156 \\
GRB 130818941 & 8.7   & 14.8 & 436 & 0.062 \\
GRB 131014513 & 2.1   & 8.5  & 741 & 0.290 \\
GRB 131108024 & 1.8   & 7.4  & 893 & 0.330 \\
GRB 140104731 & 1.5   & 15.6 & 620 & 0.172 \\
GRB 140108721 & 11.6  & 31.1 & 533 & 0.674 \\
GRB 140126815 & 14.1  & 5.0  & 498 & 0.043 \\
GRB 140304849 & 30.7  & 7.8  & 936 & 0.081 \\
GRB 140329295 & 0.6   & 65.6 & 422 & 0.179 \\
GRB 140404030 & 7.7   & 6.8  & 511 & 0.172 \\
GRB 140512814 & 11.8  & 13.4 & 378 & 0.680 \\
GRB 140621827 & 0.7   & 35.2 & 928 & 0.666 \\
GRB 140628704 & 4.9   & 4.6  & 241 & 0.201 \\
GRB 140709051 & 5.7   & 7.0  & 772 & 0.600 \\
GRB 140714268 & 27.5  & 6.1  & 807 & 0.918 \\
GRB 140716436 & 2.2   & 18.4 & 637 & 0.260 \\
GRB 140818229 & 10.2  & 10.5 & 352 & 0.852 \\
GRB 140824606 & 12.9  & 12.3 & 26  & 0.382 \\
GRB 140825328 & 3.2   & 11.3 & 353 & 0.687 \\
GRB 140917512 & 3.9   & 13.1 & 788 & 0.904 \\
GRB 141029134 & 6.9   & 15.1 & 891 & 0.009 \\
GRB 141102536 & 0.1   & 13.2 & 971 & 0.923 \\
GRB 150126868 & 13.0  & 10.3 & 420 & 0.454 \\
GRB 150127398 & 5.7   & 43.9 & 227 & 0.884 \\
GRB 150226545 & 16.2  & 5.6  & 194 & 0.542 \\
GRB 150330828 & 11.5  & 55.3 & 440 & 0.776 \\
GRB 150506398 & 27.8  & 0.6  & 96  & 0.962 \\
GRB 150508945 & 15.7  & 9.0  & 888 & 0.816 \\
GRB 150512432 & 20.2  & 10.1 & 975 & 0.579 \\
GRB 150522433 & 7.8   & 7.1  & 217 & 0.753 \\
GRB 150523396 & 19.7  & 26.4 & 648 & 0.041 \\
GRB 150702998 & 2.5   & 25.1 & 960 & 0.470 \\
GRB 150703149 & 0.0   & 0.2  & 933 & 0.193 \\
GRB 150830128 & 14.0  & 4.3  & 797 & 0.099 \\
GRB 151027166 & 40.6  & 10.7 & 440 & 0.739 \\
GRB 151030999 & 17.7  & 22.1 & 147 & 0.456 \\
GRB 160131174 & 44.0  & 5.1  & 626 & 0.282 \\
GRB 160201883 & 1.0   & 9.3  & 395 & 0.979 \\
GRB 160215773 & 44.6  & 10.6 & 352 & 0.463 \\
GRB 160219673 & 12.5  & 10.5 & 470 & 0.951 \\
GRB 160223072 & 10.5  & 9.8  & 204 & 0.737 \\
GRB 160225809 & 23.2  & 12.8 & 602 & 0.040 \\
GRB 160512199 & 9.4   & 7.4  & 558 & 0.929 \\
GRB 160519012 & 17.1  & 3.4  & 489 & 0.083 \\
GRB 160523919 & 5.4   & 10.4 & 300 & 0.861 \\
GRB 160625945 & 2.4   & 46.1 & 236 & 0.859 \\
GRB 160724444 & 1.8   & 8.4  & 720 & 0.957 \\
GRB 160821857 & 31.8  & 14.5 & 728 & 0.232 \\
GRB 160825799 & 0.6   & 10.2 & 397 & 0.096 \\
GRB 160908136 & 6.8   & 7.9  & 964 & 0.401 \\
GRB 160912521 & 5.2   & 11.3 & 904 & 0.006 \\
GRB 160919613 & 0.8   & 8.0  & 299 & 0.812 \\
GRB 161105417 & 12.7  & 13.2 & 387 & 0.021 \\
GRB 161111197 & 11.1  & 7.3  & 170 & 0.692 \\
GRB 161117066 & 77.0  & 17.5 & 316 & 0.946 \\
GRB 161119633 & 7.7   & 5.7  & 163 & 0.942 \\
GRB 170109137 & 6.4   & 7.8  & 178 & 0.786 \\
GRB 170115662 & 18.6  & 11.4 & 538 & 0.344 \\
GRB 170209048 & 8.2   & 24.6 & 562 & 0.763 \\
GRB 170302719 & 12.3  & 6.1  & 181 & 0.283 \\
GRB 170323775 & 12.7  & 8.5  & 676 & 0.496 \\
GRB 170402961 & 0.2   & 12.2 & 156 & 0.905 \\
GRB 170416583 & 12.5  & 28.6 & 317 & 0.521 \\
GRB 170514152 & 0.7   & 15.9 & 90  & 0.804 \\
GRB 170514180 & 35.9  & 15.3 & 832 & 0.904 \\
GRB 170830069 & 6.0   & 6.7  & 824 & 0.200 \\
GRB 170831179 & 6.3   & 10.4 & 307 & 0.054 \\
GRB 170923188 & 1.0   & 11.7 & 943 & 0.852 \\
GRB 171004857 & 1.4   & 7.3  & 77  & 0.908 \\
GRB 171102107 & 10.4  & 15.2 & 271 & 0.718 \\
GRB 171112868 & 9.5   & 10.2 & 675 & 0.366 \\
GRB 171120556 & 4.2   & 67.9 & 501 & 0.211 \\
GRB 171211844 & 12.4  & 15.3 & 397 & 0.036 \\
GRB 180307073 & 23.3  & 9.8  & 698 & 0.624 \\
GRB 180411519 & 26.7  & 13.0 & 642 & 0.366 \\
GRB 180416340 & 10.3  & 29.6 & 95  & 0.723 \\
GRB 180618724 & 26.2  & 5.4  & 757 & 0.699 \\
GRB 180620354 & 5.9   & 9.4  & 265 & 0.886 \\
GRB 180710062 & 13.5  & 4.3  & 802 & 0.556 \\
GRB 180720598 & 10.0  & 7.1  & 136 & 0.182 \\
GRB 180728728 & 10.0  & 11.8 & 512 & 0.801 \\
GRB 180822423 & 2.8   & 18.5 & 415 & 0.558 \\
GRB 180906988 & 1.0   & 13.7 & 842 & 0.088 \\
GRB 180929453 & 0.6   & 9.6  & 725 & 0.494 \\
GRB 181008877 & 27.9  & 6.9  & 630 & 0.909 \\
GRB 181119606 & 1.8   & 49.5 & 232 & 0.622 \\
GRB 181122381 & 0.3   & 18.4 & 933 & 0.468 \\
GRB 181203880 & 0.9   & 9.1  & 821 & 0.389 \\
GRB 181222279 & 40.8  & 28.9 & 155 & 0.790 \\
GRB 190114873 & 1.5   & 6.4  & 361 & 0.182 \\
GRB 190228973 & 8.0   & 10.5 & 947 & 0.570 \\
GRB 190310398 & 4.1   & 17.1 & 258 & 0.968 \\
GRB 190326314 & 2.1   & 18.8 & 685 & 0.691 \\
GRB 190610750 & 1.2   & 20.5 & 835 & 0.549 \\
GRB 190611950 & 20.1  & 44.2 & 920 & 0.610 \\
GRB 190719624 & 1.6   & 27.5 & 751 & 0.714 \\
GRB 190806675 & 1.2   & 16.7 & 627 & 0.184 \\
GRB 190828542 & 38.5  & 2.2  & 979 & 0.927 \\
GRB 190829830 & 5.6   & 21.9 & 767 & 0.791 \\
GRB 190901890 & 20.0  & 1.1  & 608 & 0.620 \\
GRB 190930400 & 40.3  & 7.1  & 832 & 0.259 \\
GRB 191019970 & 29.8  & 28.8 & 768 & 0.077 \\
GRB 191026350 & 4.1   & 6.0  & 896 & 0.050 \\
GRB 191031025 & 10.4  & 6.9  & 774 & 0.133 \\
GRB 191101895 & 1.9   & 6.8  & 396 & 0.027 \\
GRB 191111364 & 16.4  & 6.5  & 647 & 0.133 \\

\end{longtable*}

\end{document}